\documentclass[twocolumn,showpacs,preprintnumbers,amsmath,amssymb]{revtex4}
\usepackage{graphicx}% Include figure files
\usepackage{dcolumn}% Align table columns on decimal point
\usepackage{bm}% bold math

\begin{document}

\preprint{APS/123-QED}

\title{Vibrational dynamics of confined granular materials}

\author{Emilien Az\'ema, Farhang Radja\"\i, Robert Peyroux, Fr\'ed\'eric Dubois }
\affiliation{LMGC, CNRS - Universit\'e Montpellier II, Place Eug\`ene Bataillon, 34095 Montpellier cedex 05, France.}
\email{azema@lmgc.univ-montp2.fr}

 \author{Gilles Saussine}
\affiliation{Physics of Railway Systems, Innovation and Research Departement of SNCF, 45 rue de Londres, 75379 PARIS Cedex 08}

\date{\today}

\begin{abstract}
By means of two-dimensional contact dynamics simulations, we analyze the 
vibrational dynamics of a confined granular layer in response to  
harmonic forcing.  
We use irregular polygonal grains allowing for strong variability of solid fraction. 
The system involves  a jammed state separating passive (loading) and active (unloading) states. 
We show that an approximate expression of the  packing resistance force as a function of the displacement of the free retaining wall from the jamming position 
provides a good description of the dynamics. 
We study in detail the scaling of displacements and velocities with 
loading parameters. In particular, we find that, for a wide range of 
frequencies, the data collapse by scaling the displacements with the 
inverse square of frequency, the inverse of the force amplitude and the square of gravity. 
Interestingly, compaction occurs during the extension of the packing, followed by  decompaction in the contraction phase. We show that the mean compaction rate increases linearly with frequency up to a characteristic frequency and then it declines  in inverse proportion to frequency. 
The characteristic frequency is interpreted in terms of the time required for the relaxation of the packing  through collective grain rearrangements between two equilibrium states. 
\end{abstract}

\maketitle

\section{Introduction}
Depending on the  frequency and amplitude of accelerations, vibrated 
granular materials give rise to various phenomena such as compaction, (\cite{Knight94,Sano2005})
convective flow(\cite{aoki96,liffman97,Knight93}), size segregation and standing wave patterns at the free surface (\cite{aoki96,aoki96b,Clement96}).    
Particle rearrangements induced by vibrations lead to lower shear strength and larger 
flowability. In the full fluidization regime, there are no permanent 
contacts between particles and the system behaves as a dissipative gas \cite{jaeger96b}. 
Particle bed reactors are sometimes fluidized by this method instead of 
upward gas flow (\cite{Brennen93}). When particle accelerations remain below the gravitational 
acceleration, the system keeps its static nature and the vibrational 
energy propagates through a rather compact network of interparticle contacts. 
This leads to enhanced bulk flow in hoppers and chutes \cite{weathers97,wassgren97c}. 

On the other hand, vibrations at high frequency 
and low amplitude lead to slow (logarithmic) decay of the pore space 
as a function of time \cite{Knight93}. 
Efficient vibro-compaction of dry and wet granular materials is a crucial issue 
in numerous applications such as the casting of fresh concrete. 
The tamping operation on railway ballast  is another  example where the 
vibrations of tamping bars are 
used to restore the initial geometry of the track distorted as a result of 
ballast settlement (\cite{SAUSSINE2004,XIMENA2001,MORGAN81}). 
The maintenance cost becomes crucial
with the increase of commercial speed.

We may distinguish two methods for inducing vibrational dynamics: 
1) by imposed cyclic displacements of a wall or the container (shaking); 
2) by cyclic modulation of a confining stress. 
The first method has been used in most experiments on granular beds  
(\cite{Luding95,ben-naim96,ben-naim97,Hunt99,Kudrolli2004,Josserand2000}). 
In this case, the control parameters are the amplitude $a$ and the frequency $\nu$ of the vibrations corresponding to a maximal acceleration $a\omega^2$ where $\omega=2\pi \nu$.
When a material is moulded inside a closed box, the vibrations should rather 
be induced by varying a confining force, e.g. a force acting on a wall. 
Then, the amplitude of displacements is a function of the forcing frequency, 
and the level of particle accelerations 
depends on both the applied cyclic force and the reaction force of the packing. 
In any case, an efficient compaction process  
requires periods of release of the packing so that the grains 
can move with respect to their neighbors.

In this paper, we explore such a system where a harmonic 
force $f$ is exerted on a lateral wall of a closed box, 
all other walls remaining immobile. 
The force $f$ is varied between zero and a maximum value $f_{max}$. 
During a period, $f$ is large enough to equilibrate the packing reaction force  
except for a short laps of time when $f$  declines to zero. 
Then, the packing can flow under the action of its own weight, pushing the retaining wall away. 
We are interested here in the evolution of the packing in the course of 
harmonic loading and its scaling  with loading parameters (frequency, force maximum). 

We used numerical simulations by the contact dynamics approach  
as a discrete element method (DEM) in a two-dimensional geometry with 
a small number of particles (\cite{jean92,mo04a}).
Each simulation is repeated for several independent configurations and the 
results are analyzed in terms of ensemble average behaviors. 
The particles are rigid and polygon-shaped.
We focus on the displacements of the free retaining wall and the 
compaction of the packing. 
Most results presented below concern the short-time behavior 
where the solid fraction increases linearly with time.
The frequency is varied from 1 to 60 Hz and its influence is analyzed 
by considering characteristic times involved in the loading and 
unloading intervals of time.
We first introduce the numerical procedures. Then, we present the main 
findings concerning the passive and active dynamics, the evolution of the solid 
fraction and scaling with the loading parameters. 

\section{Numerical procedures}
The simulations were carried out by means of the contact dynamics (CD) method with 
irregular polygonal particles (\cite{jean92,mo04a}). 
The CD method is based on implicit time integration of the equations 
of motion and a nonsmooth formulation of  
mutual exclusion and dry friction between particles.  
This method requires no elastic repulsive potential and no smoothing 
of the Coulomb friction law 
for the determination of forces. For this reason, the simulations can be 
performed with large time steps compared to molecular dynamics simulations.  
We used LMGC90 which is a multipurpose software  
developed in our laboratory, capable of modeling a collection of deformable or undeformable particles of
various shapes by different algorithms (\cite{DUBOIS2003}).

%----------------------------------------------------

\begin{figure}
\includegraphics[width=6cm]{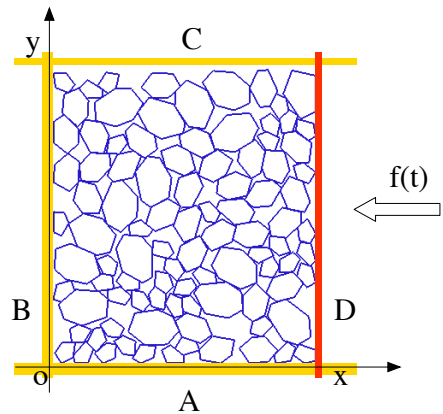}
\caption{ The geometry of the packing. \label{fig1}}
\end{figure}

The samples are composed of irregular pentagons, hexagons, 
and heptagons of three different diameters:  
50\% of diameter $d_{min}=2.5$ cm, 34\% of diameter $3.75$ cm, 
16\% of diameter $d_{max}=5$ cm; see fig. \ref{fig1}. 
The particles are initially placed on a square network in a rectangular 
box and compressed by downward motion of the upper wall 
(wall C in fig. \ref{fig1}) at zero gravity. Then, the gravity is set to $g$ and the 
upper wall is raised 1 cm and fixed. 
The right wall (wall D in fig. \ref{fig1}) is allowed to move 
horizontally (x direction) and subjected to a driving force: 
\begin{equation}
f(t)=\frac{(f_{max}+f_{min})}{2}-\frac{(f_{max}-f_{min})}{2}\sin \omega t, 
\end{equation}
where $f_{max}$ and $f_{min}$ are the largest and lowest compressive 
(positive) forces acting on the wall. 

If $f_{min}$ is above the (gravitational) force 
exerted by the grains on the free wall, $f$ will be large enough 
to prevent the wall from backward motion during the whole 
cycle.  
In other words, the granular material is in "passive state"  
in the sense of Rankine's states and the 
major principal stress direction is horizontal (\cite{nedderman92}). In this limit,
no extension will occur following the initial contraction. 
On the other hand, if $f_{max}$ is below the force exerted by the grains, 
$f$ will never be large enough to prevent the extension of the packing. 
This corresponds to  the "active state" where the major 
principal stress direction remains vertical. 
In all other cases, both contraction and extension occur during each period, 
and the displacement $\Delta x$ of the free wall will be controlled by  $f_{min}$.
In the simulations reported below, we set $f_{min}=0$. 
This ensures the largest possible displacement of the wall in the active state. 
We used four different values of $f_{max}$ ranging from $5\;.10^3$ N to $2\;.10^4$. 

The simulations were carried out with $N_p=95$ grains in the box and each 
simulation was repeated with seven independent grain configurations. 
The mean behavior for each set of parameters is obtained 
by ensemble averaging over seven independent data sets.  
Larger samples can be simulated, but that  requires much more 
computational effort  for a parametric study over many cycles.
Thus, our system represents rather a thin granular layer. 
The coefficient of friction between the grains and with the horizontal walls 
was fixed to 0.4, but it was 0 at the vertical walls. 
With a time step equal to $2.5\; 10^{-4}$ s 
we could perform high-quality simulations in which the largest cumulative error 
on grain positions was bellow 1\%. 

% ---------------------------------------------------------------------
% Dynamics

\section{Active and passive dynamics}

\begin{figure}
\includegraphics[width=6cm]{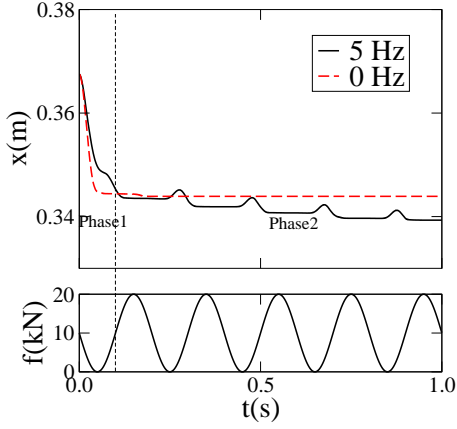}
\caption{ The evolution of the displacement $x$ of the free wall (up) in response to 
harmonic loading (down). \label{fig2}}
\end{figure}

We first consider the motion $x(t)$ of the free wall (wall D in Fig. \ref{fig1}) 
which reflects the dynamics of the grains in the cell in 
response to harmonic forcing.
Figure \ref{fig2} shows $x(t)$ (averaged over seven independent simulations) 
for frequency $\nu=5$ Hz over a time interval $\Delta t = 1$ s. 
We distinguish a fast initial contraction ($t<0.1$ s) followed by slow 
contraction (decreasing $x$) over four periods. 
The initial contraction is a consequence of the gap left 
between the free surface of the packing and the upper wall. 
This initial volume change is almost independent of frequency. 
The subsequent periodic motion of the wall takes place around this 
confined state and will be at the focus of this paper.

\begin{figure}
\includegraphics[width=6cm]{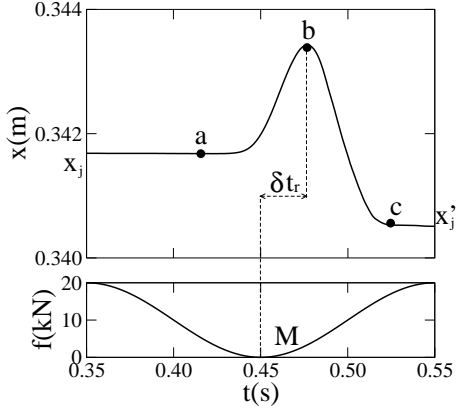}
\caption{ A zoom on a single period; see Fig. \ref{fig2}. \label{fig3}}
\end{figure}

A zoom on a single period is shown in Fig. \ref{fig3}.
The period begins at the jamming position $x=x_j$ corresponding to 
the jamming position reached in the preceding period. 
The motion of the wall begins (point a in Fig. \ref{fig3}) only when the 
applied force $f$ declines near to its minimum $f_{min}=0$. 
The maximum displacement $\Delta x_{max}$ occurs at a later time $\delta t_r$ (point b). 
From a to b, the  force exerted by the packing on the free wall is above 
the applied force, so that the wall moves backward (extension). 
In this phase, the packing is in an active state. 
The inverse situation prevails from b to c where the grains are pushed towards the box (contraction). 
Then, the packing is in a passive state. 
The new jamming position $x'_j$ is below the jamming position $x_j$ 
reached at the end of the preceding period. 
The difference $x_j - x'_j$ represents the net compaction of the packing 
over one period.
For a given frequency $\nu$, the phase difference $\delta t_r$ is the same for all periods.
The displacement amplitude $\Delta x_{max}$ is a 
function of $f_{max}$ and $\nu$, as we shall see below.

\begin{figure}
\includegraphics[width=6cm]{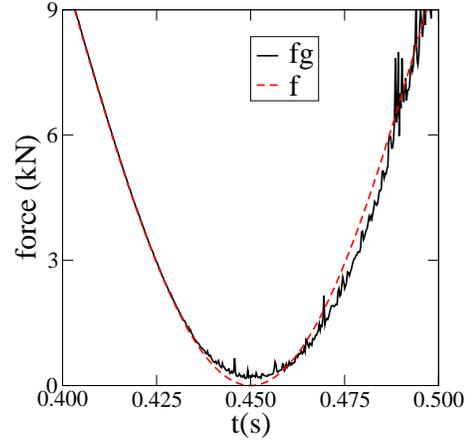}
\caption{ The force $f_g$ exerted by the grains and the driving force $f$  on the free wall as 
a function of time $t$. \label{fig4}}
\end{figure}

The motion of the free wall is governed by the equation of dynamics, 
\begin{equation}
f-f_g=m\ddot{x},
\label{eq2}
\end{equation}
where $f_g$ is the horizontal force exerted by the packing on 
the wall (fig \ref{fig1}). Figure \ref{fig4} displays $f_g$ as the function of time 
for $f_{max} = 2\;10^4$ N. We see that $f_g$ follows 
closely the variations of $f$. In particular, in the jammed state we 
have $f=f_g$ so that $\ddot x =0$ in this state. 
This means that, in its most general form, $f_g$ is a function of $f$. 

\begin{figure}
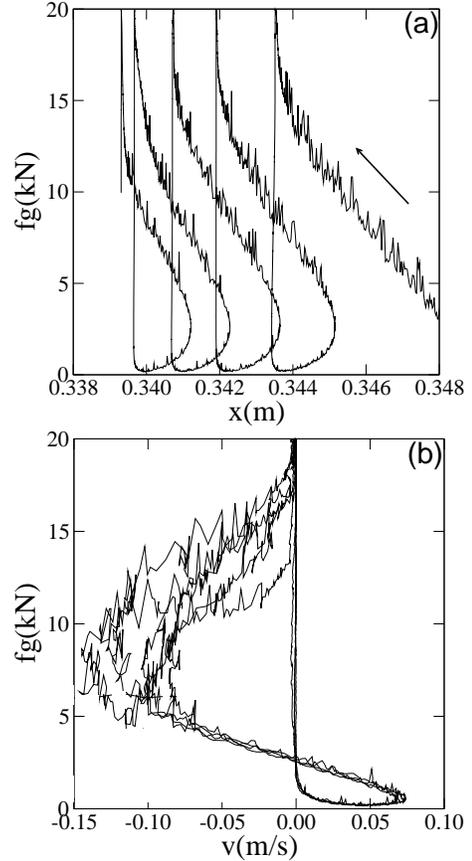

\includegraphics[width=6cm]{Fig/fig5a.eps}
\includegraphics[width=6cm]{Fig/fig5b.eps}
\caption{Force $f_g$ exerted by the grains on the free wall 
as a function of displacement $x$ (a) and the velocity $v$ (b). (\label{fig5}}
\end{figure}

Figure \ref{fig5}(a) shows $f_g$ vs. $x$ over four periods. 
In the active phase, $f_g$ grows slightly with $x$.  
In the passive phase, it grows faster and almost linearly as $x$ decreases. 
The vertical line corresponds to the jammed state 
where $f_g$ decreases with $f$ at $x=x_j$. 
We also clearly observe in Fig. \ref{fig5}(a)  two transients :
1) unjamming and the onset of the active state, 
2) jamming from the passive state. 
It is remarkable that, although $x_j$ decreases at the end of  
each period, the dynamics remains self-similar up to a translation along 
displacement coordinates.   

Figure \ref{fig5}(b) displays $f_g$ as a function of the velocity $v \equiv \dot x$. 
We again observe the passive ($ v<0$) and active ($v>0$) states together with the 
jamming and unjamming transients before and after the jammed state 
($v=0$ and $x=x_j$). 
The data from all periods follow the same variations 
except for the jamming transient where a slight decrease of 
the maximum negative velocity $v_{max}$ can be noticed in each period.

\begin{figure}
\includegraphics[width=6cm]{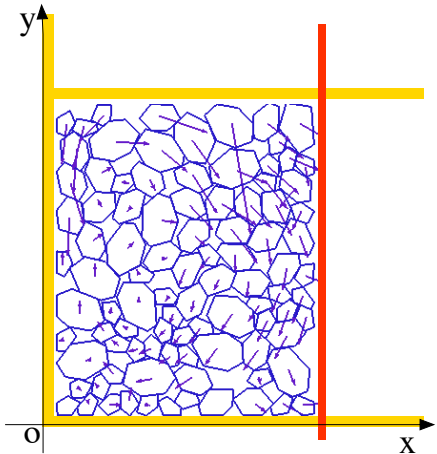}
\caption{ Particle displacements over one period. \label{fig6}}
\end{figure}

\begin{figure}
\includegraphics[width=6cm]{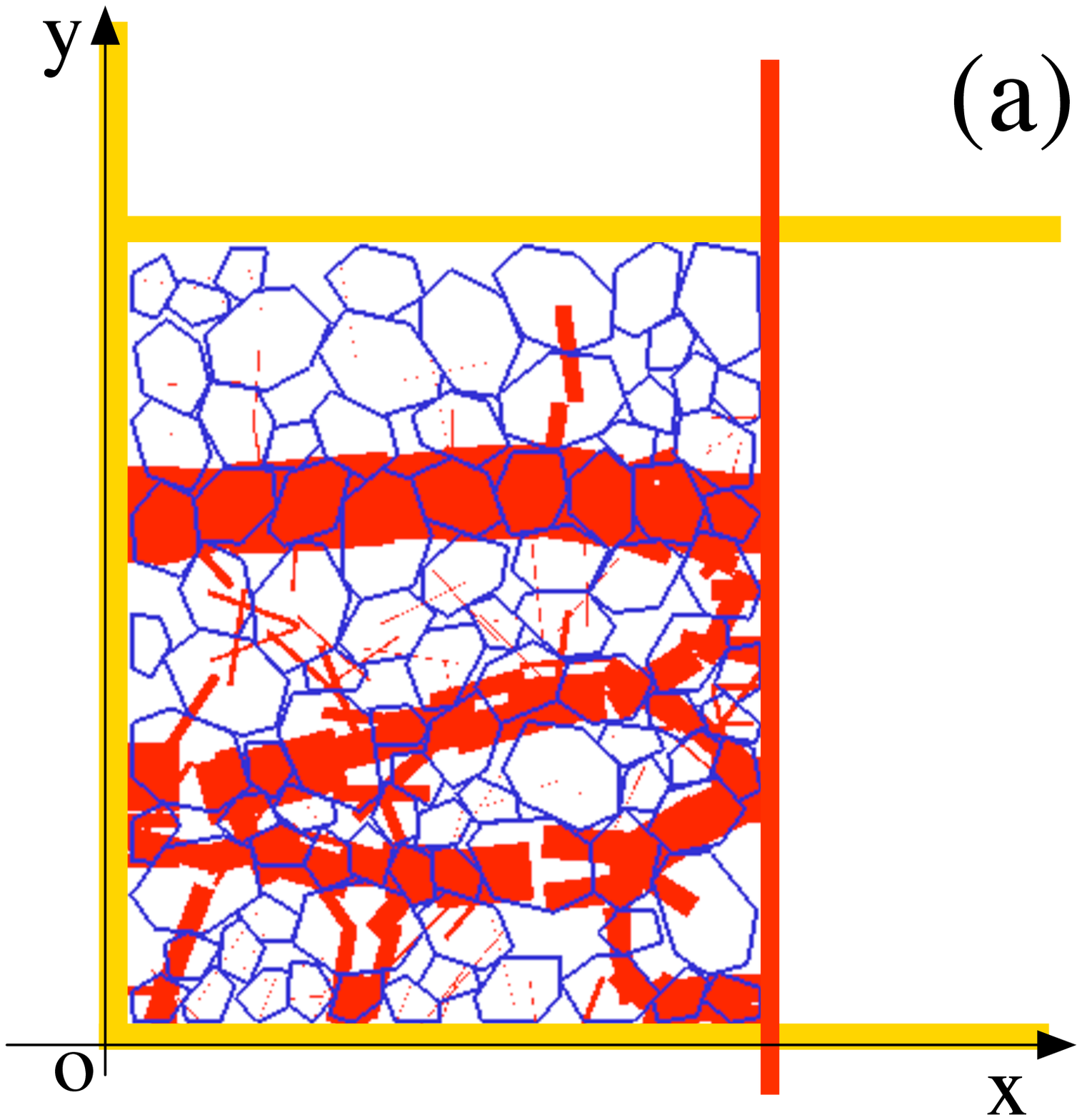}
\includegraphics[width=6cm]{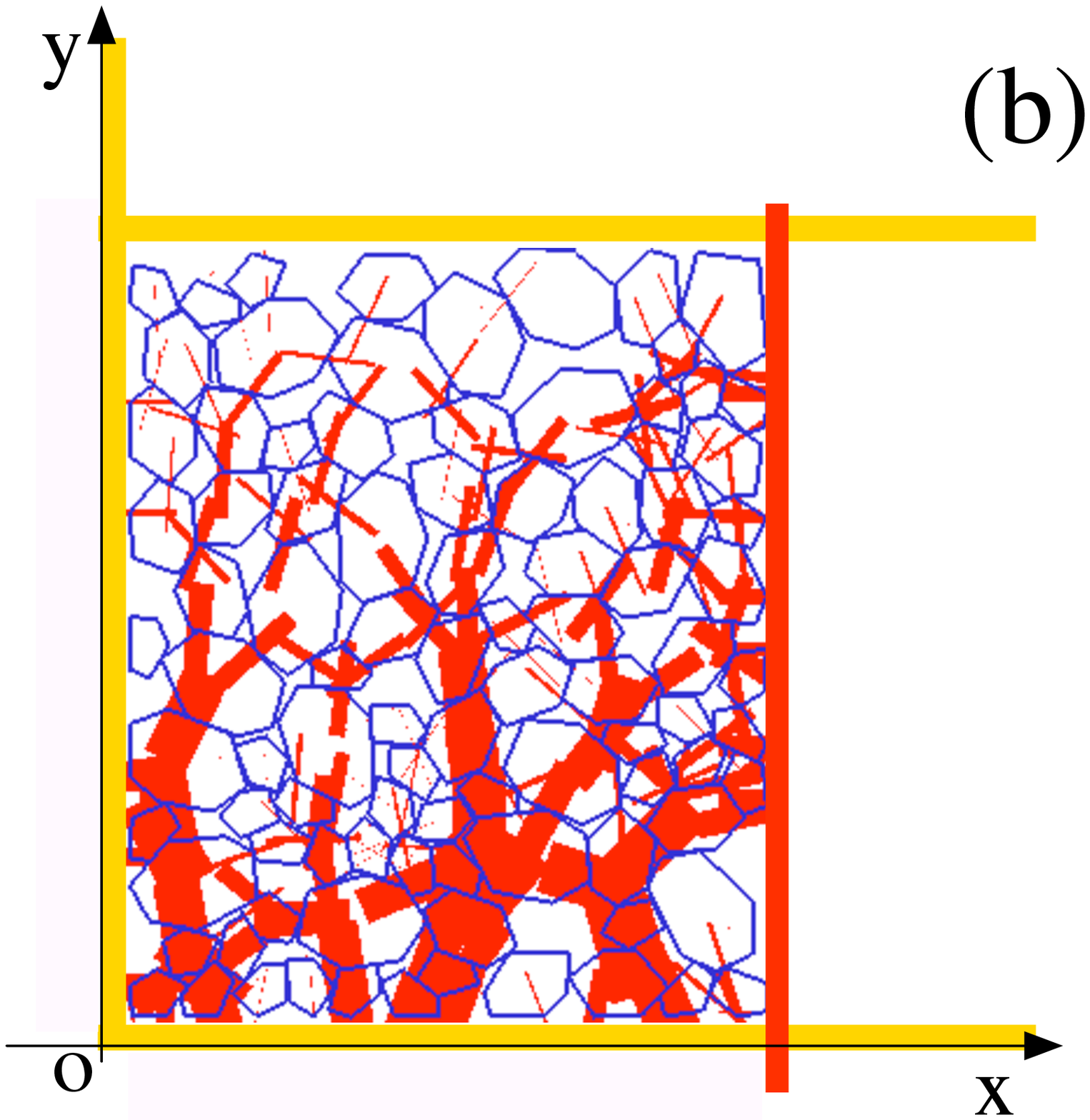}
\caption{ Normal forces in the passive (a) and active (b) states. 
Line thickness is proportional to the force. \label{fig7}}
\end{figure}

Although we focus here on the average dynamics of the packing, 
i.e. the displacements of the free wall, it is important to note that 
the grain velocity field is not a simple oscillation around an average position. 
The grains undergo a clockwise convective motion in the cell as shown in Fig. \ref{fig6}.  
On the other hand, the contact forces evolve between a fully jammed state, 
where nearly horizontal force chains dominate (Fig. \ref{fig7}(a)), and the active state, 
where nearly vertical gravity-induced chains can be observed (Fig.  \ref{fig7}(b)).

% ----------------------------------------------------------------------------------
% Model

\section{A phenomenological model}

\begin{figure}
\includegraphics[width=6cm]{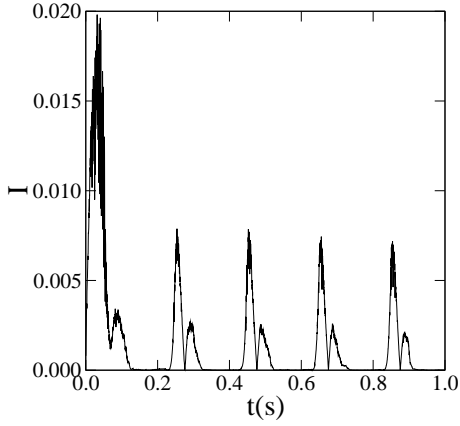}
\caption{The evolution of the inertia number $I$ over five periods. \label{fig8}}
\end{figure}

To predict the motion of the free wall from equation \ref{eq2}, 
we need to express the force $f_g$ as a function of $x$ and $v$. 
It is obvious that in the jammed state at $x=x_j$, the force $f_g$ is 
a reaction force balancing exactly the driving force $f$ so that $v=0$. 
On the other hand, the inertia effects are small compared to static forces. To show this, 
we may use  a dimensionless number $I$ defined by (\cite{GDRMIDI2004}): 
\begin{equation}
I=\dot \varepsilon \sqrt{\frac{m}{p}},
\label{eq3}
\end{equation}
where $\dot \varepsilon=\dot x /x$ is the deformation rate, 
$m$ is the total mass, and $p$ is the average pressure.
The evolution of $I$ is plotted as a function of time in Fig. \ref{fig8}. 
The two peaks in each period correspond to the maxima of the largest 
velocities in absolute value shown in Fig. \ref{fig5}(b).  
We see that $I < 0.02$, implying that  $f_g$ should not depend crucially on $v$. 
Let us note  that the plot of $f_g$ vs $v$ in Fig.\ref{fig5}(b) 
does not represent the {\em explicit} dependence of $f_g$ on $v$; 
it is a consequence of the equation of dynamics and, as we shall see below, it 
can be reproduced by assuming no dependence of  $f_g$ on $v$. 

\begin{figure}
\includegraphics[width=6cm]{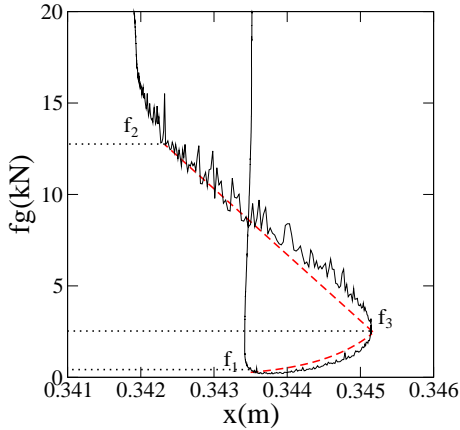}
\caption{ Variation of the packing reaction force $f_g$ with displacement 
$x$ over one period (full line) and an approximate fitting form (dashed line). \label{fig9}}
\end{figure}

We now introduce a simple phenomenological model in which 
the expression of $f_g$ as a function of $x$ is extracted from the numerical 
data plotted in Fig. \ref{fig5}(a). 
As shown in Fig. \ref{fig9}, two distinct fitting forms are to be considered 
for the active and passive states.
Ignoring the  jamming and unjamming short transients, an exponential 
form provides a nice fit for the active branch whereas a linear fit 
seems fairly good in the  passive state. 
Hence, with a good approximation we can write
\begin{equation}
f_g= 
\left\{
\begin{array}{lll}
\alpha + \beta  \ e^{k \ (x-x_j) } & \mbox{active}, &  \\
\alpha' + \beta' \  \{1 + k' \ (x-x_j) \}  & \mbox{passive} &  
\end{array}
\right.
\label{eq:model}
\end{equation}
with 
\begin{equation}
\begin{array}{lll}
\alpha  &=&     \left(f_3 - f_1 e^{k \Delta x_{max}}\right) /  
 \left(1 - e^{k \Delta x_{max}}\right), \\
\beta   &=&   \left( f_1 - f_3 \right) /  
 \left(1 - e^{k \Delta x_{max}}\right),     \\
\alpha'  &=&      \left(f_2 (1 +  k' \Delta x_{max}) - f_3   \right) /  
 \left(   k'   \Delta x_{max}  \right),   \\
\beta'   &=&      \left( f_3- f_2  \right) / 
 \left(   k'   \Delta x_{max}  \right).
\end{array}
\label{eq:para}
\end{equation}
The constant forces $f_1$, $f_2$ and $f_3$ correspond to the values of $f_g$ at 
the unjamming transient, the jamming transient and the point of transition from active to 
passive states, respectively (see Fig. \ref{fig9}).  
Clearly, because of the action of gravity and  jamming transition, 
we have $f_1 > f_{min}$ and $f_2 < f_{max}$. 

\begin{figure}
\includegraphics[width=6cm]{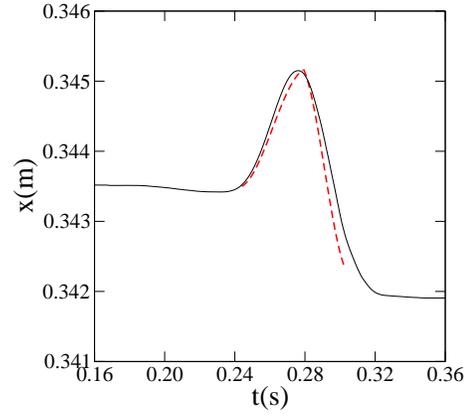}
\caption{Displacement $x$ of the free wall as a function of time (full line) and 
analytical fit from the phenomenological model (dashed line). \label{fig10}}
\end{figure}
 
We substitute this expression \ref{eq:model} in equation \ref{eq2} 
and we solve for $x$. 
Analytical solution can be obtained for the passive linear part. 
An approximate solution can be given also for the active part by 
expending the exponential function to leading order.  
Figure \ref{fig10} shows the evolution of the position $x$ 
for one period together with the solution of the model.

\begin{figure}
\includegraphics[width=6cm]{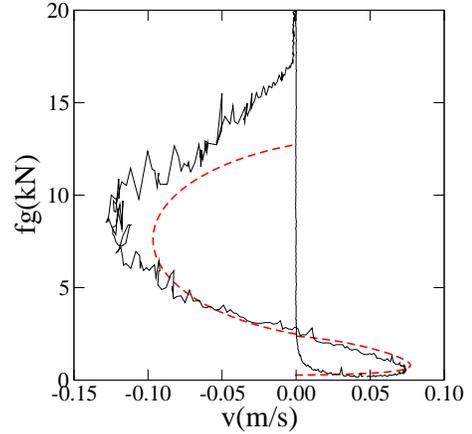}
\caption{Force $f_g$ exerted by the grains on the free wall versus 
velocity $v$ over one period (full line) and analytical fit from 
the phenomenological model (dashed line).\label{fig11}}
\end{figure}

\begin{figure}
\includegraphics[width=6cm]{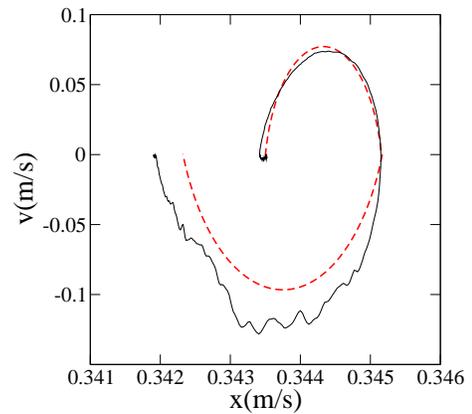}
\caption{Phase space trajectory over one period  (full line) and 
analytical fit from the phenomenological model (dashed line).\label{fig12}}
\end{figure}

The parameters $k$, $k'$ are adjusted in order to get the best fit for the plot. 
The  continuity of the fit at transition between passive and active states is ensured by 
the very choice of the coefficients according to Eq. \ref{eq:para}.  
Although we did not take into account the transients, the analytical plot fits correctly the data. 

Figure \ref{fig11} displays $f_g$ vs $v$ for one period, together with the analytical fit obtained as 
solution to  
Eq. \ref{eq2} given the expression \ref{eq:model} of $f_g$ as a function of $x$.   
Again, excluding jamming and unjamming  transients, the analytical solution provides a fairly 
good approximation for the simulation data. 
Fig. \ref{fig12} shows the trajectory of the motion in the phase space $(\dot x,x)$ for 
one period, both from direct data and the model. 
The fit is globally acceptable  although the velocity is  under-estimated  in the passive state. 

The model parameters $k$ and $k'$ remain nearly the same over all periods.  
This means that the dynamics at short times 
($\Delta t < 1$ s) is weakly dependent on the solid fraction. The  
parameters $k$ and $k'$ change, however, with loading parameters ($\nu$, $f_{max}$, etc) 
unless the displacements and the forces $f_1$, $f_2$ and $f_3$ are scaled with these parameters. 
This point will be discussed in detail below.

% ------------------------------------------------------------------------------
% Compaction

\section{Compaction}

In order to evaluate the solid fraction $\rho$, we consider a control volume enclosing 
a portion of the packing inside the simulation cell. 
This volume does not include the initial gap between the top 
of the packing and the upper wall. 
The initial value of the solid fraction is $0.75$ and, since the grains are angular-shaped, 
its variations $\Delta \rho$ from the initial state are large.   

\begin{figure}
\includegraphics[width=6cm]{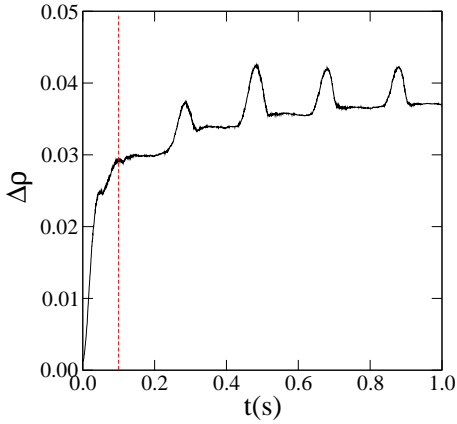}
\caption{Evolution of the solid fraction $\Delta\rho$ from the initial state 
as a function of time over several periods. \label{fig13}}
\end{figure}

Figure \ref{fig13} shows the evolution of $\Delta\rho$  for several periods. 
We observe an initial compaction of $3\% $ occurring in $0.1$ s. 
The subsequent evolution of the solid fraction takes place 
in a more compact state with a small increase in each period. 

We use $\rho_0 = 0.77$ at the end of the first period as the reference 
value for solid fraction. The relative compaction of the packing is given 
by $\Delta\rho / \rho_0$. The compaction rate $\eta$ over several periods 
and for a total time interval $\Delta t$ is 
\begin{equation}
 \eta \equiv \frac{1}{\rho_0} \frac{\Delta\rho}{\Delta t}.
 \label{eq:eta} 
\end{equation}

\begin{figure}
\includegraphics[width=6cm]{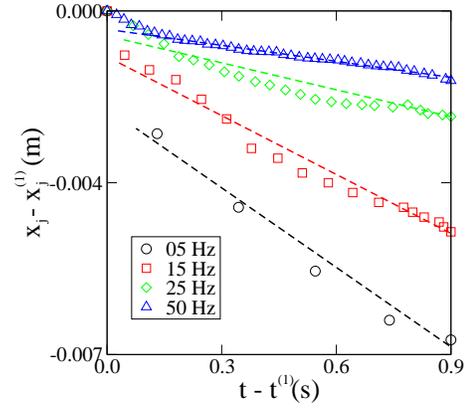}
\caption{Evolution of the jamming position $x_j$ 
from the position $x^{(1)}$ reached at $t=t^{(1)} =  0.1$ s for four 
different frequencies. \label{fig14}}
\end{figure}

Fig. \ref{fig14} shows the jamming 
position $x_j $ as a function of time for different frequencies for $\Delta t < 1$ s. 
At such short times,  it can be assumed, with a good approximation, that 
the solid fraction declines linearly in time. Generally, the behavior 
slows down logarithmically  at  longer times \cite{ben-naim96}. 
This means that at short times, in which we are interested in this paper, 
the compaction rate is nearly constant, and we have  
\begin{equation}
 \eta  = \frac{\Delta\rho_1}{\rho_0} \ \nu, 
 \label{eq:eta1} 
\end{equation}
where $\Delta\rho_1$ is the compaction per period. 
For $\nu = 5$ Hz and $f_{max} = 2\;10^4$ N, we have 
$\eta \simeq 0.009$  s$^{-1}$.   

\begin{figure}
\includegraphics[width=6cm]{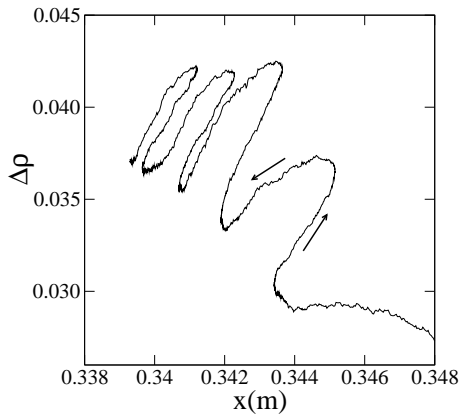}
\caption{Variation  $\Delta\rho$ of the solid fraction from the initial state 
as a function of the displacement $x$ of the free wall. \label{fig15}}
\end{figure}

Interestingly, compaction occurs in the active state, i.e. during the 
extension of the packing, and not during contraction! 
This is shown in Fig. \ref{fig15}, where the variation  $\Delta \rho$ of the solid fraction is plotted as a function of $x$. The solid fraction increases during extension (increasing $x$) 
and decreases during contraction (decreasing $x$). 

Compaction upon the reversal of the direction of shearing  is a well-known 
property of granular media \cite{mitchell05}. 
Low-amplitude cyclic shearing leads to cumulative compaction of a granular material.  
At larger amplitudes, the compaction is followed by decompaction (dilation) 
and no net compaction can be observed over a full cycle.   
The situation is slightly different in our system in the presence of a jammed state.      
Compaction is a consequence of unjamming and it is pursued during the whole active state. Decompaction takes place in the passive state, but it is cut short by fast jamming.  
The outcome of a full cycle is thus a net compaction of the packing. 

% ------------------------------------------------------------------------------------
% Scaling

\section{Scaling with loading parameters}

\begin{figure}
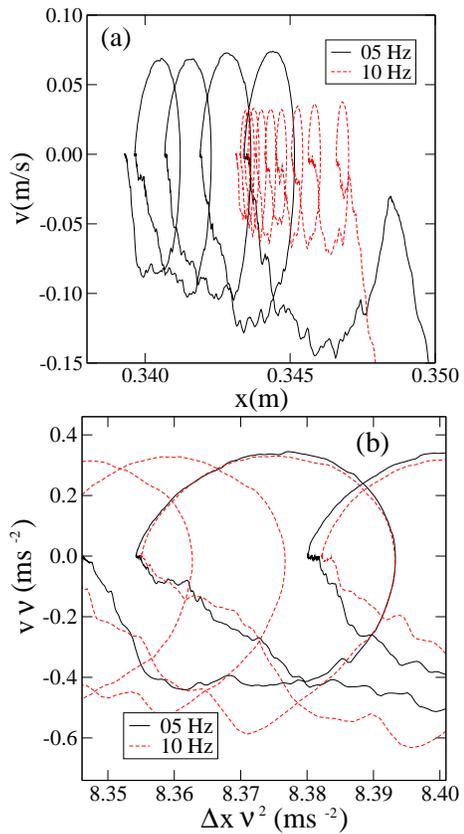

\includegraphics[width=6cm]{Fig/fig16a.eps}
\includegraphics[width=6cm]{Fig/fig16b.eps}
\caption{Phase space trajectories for two frequencies without scaling (a) 
and with scaling (b) of the displacements and velocities. \label{fig16}}
\end{figure}

In the last three sections, we analyzed the vibrational dynamics and compaction  for a 
single frequency $\nu = 5$ Hz. 
Similar simulations were performed for several frequencies 
ranging from $1$ Hz to $60$ Hz. 
Up to a change in time and length scales, all simulations yield similar results both for dynamics and compaction independently of the applied frequency. 
This can be seen, for example, in Fig. \ref{fig16}(a) where the phase 
space trajectory is shown for $\nu = 5$ Hz and $\nu = 10$ Hz. 
Fig. \ref{fig16}(b) shows that the data from both simulations collapse nicely on 
the same curve by simply scaling the 
displacements $\Delta x$ by $\nu^{-2}$ and the velocities $v$ by  $\nu^{-1}$.  

\begin{figure}
\includegraphics[width=6cm]{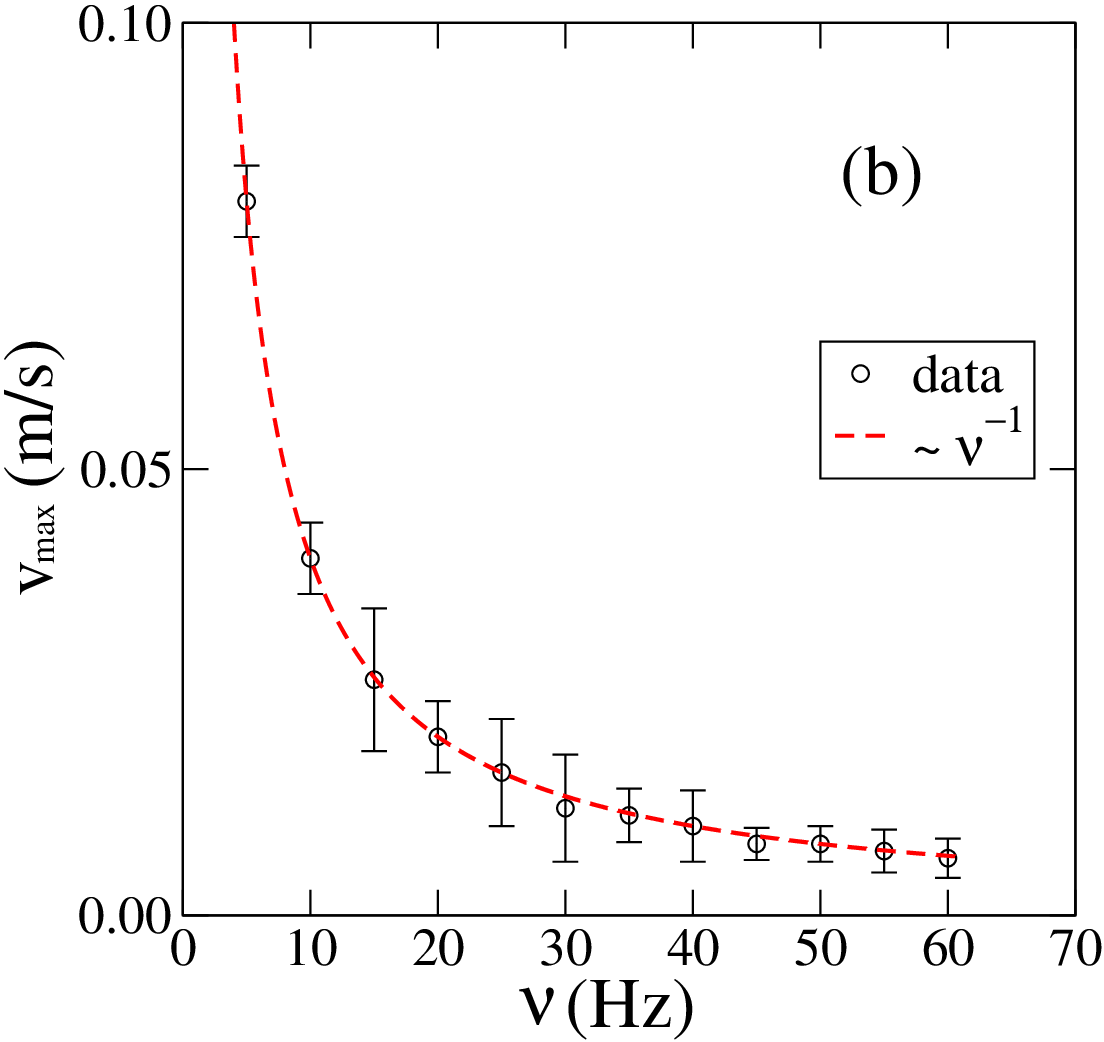}
\includegraphics[width=6cm]{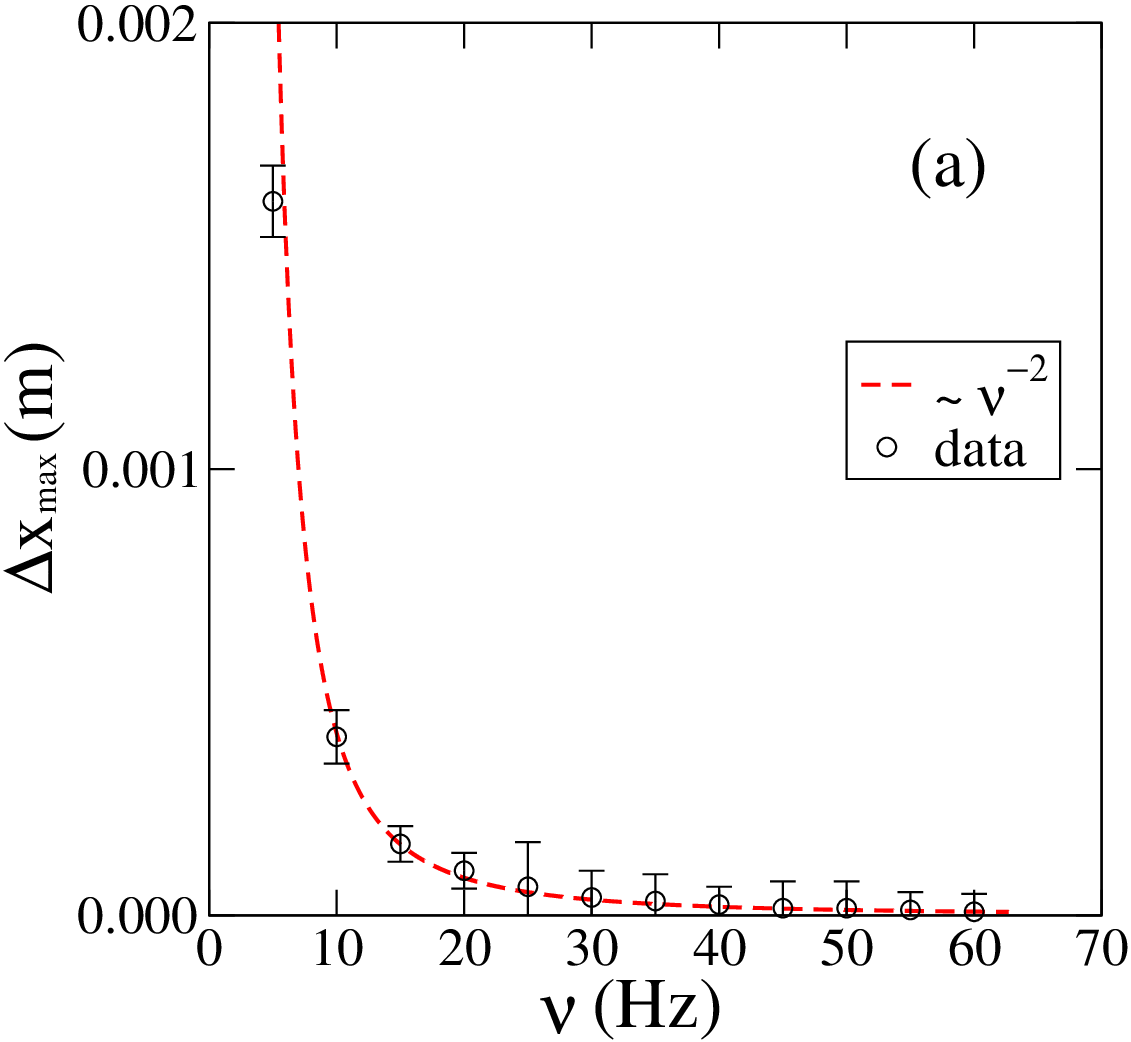}
\caption{ Maximum 
displacement ${\Delta x}_{max}$ (a) and the maximum velocity $v_{max}$ (b) 
as a function of the frequency $\nu$. \label{fig17}}
\end{figure}

This scaling is suggested by a dimensional analysis of the average dynamics of the packing. 
The frequency $\nu$ sets the time scale $\tau = \nu^{-1}$. 
Force scales are set by the largest driving force $f_{max}$ in the passive state and 
the grain weights $mg$ as well as the smallest driving force $f_{min}$ in the active state. 
Hence, dimensionally, for fixed values of  $mg$, $f_{min}$ and $f_{max}$,  
all displacements are expected to scale with $\nu^{-2}$ and 
all velocities with $\nu^{-1}$. 
To check directly this scaling, in Fig. \ref{fig17} we have plotted the maximum 
displacement ${\Delta x}_{max}$ in the active state and the maximum velocity $v_{max}$ in 
the passive state as a function of $\nu$. 
The corresponding fits by $\nu^{-2}$ and $\nu^{-1}$ are excellent. 

\begin{figure}
\includegraphics[width=6cm]{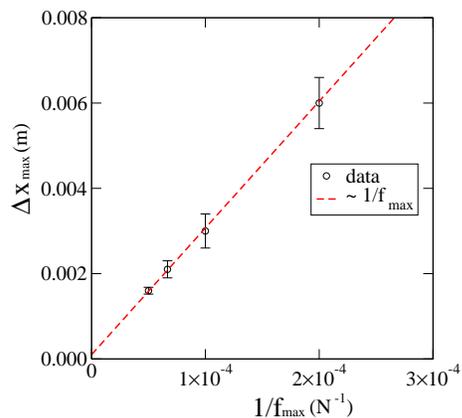}
\caption{Scaling of the maximum displacement ${\Delta x}_{max}$ with the 
force amplitude $f_{max}$. \label{fig18}}
\end{figure}

\begin{figure}
\includegraphics[width=6cm]{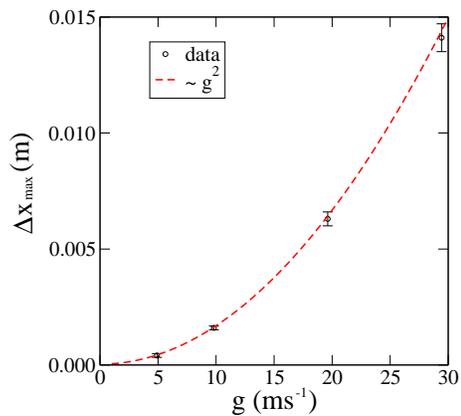}
\caption{Scaling of the maximum displacement ${\Delta x}_{max}$ with 
gravity $g$. \label{fig19}}
\end{figure}

The influence of loading force parameters $mg$, $f_{min}$ and $f_{max}$ should be 
analyzed separately for each regime. In the passive state, 
$f_{max}$ is the dominant force and it is exactly balanced by 
$f_g$  in jamming transition. 
On the other hand, in the active state,  $mg$ is the dominant force as  
$f$ remains small compared to $mg$ in this state. 
The maximum displacement  ${\Delta x}_{max}$ at transition from active 
to passive state is determined in a subtle way by both $f_{max}$ and $mg$. 
If gravity were the only driving force in the active state, ${\Delta x}_{max}$ 
would simply scale with $g \nu^{-2}$ independently of $f_{max}$. 
However, our data show that  ${\Delta x}_{max}$  
varies as $f_{max}^{-1}$; Fig. \ref{fig18}. 
A plausible dimensional interpretation is to assume 
that ${\Delta x}_{max}$ is controlled by the 
ratio $mg / f_{max}$ representing the relative importance of the gravitational  
to loading forces. Then, the following simple expression can be proposed for the 
scaling with loading forces:    
\begin{equation}
 {\Delta x}_{max}  = C \left( \frac{mg}{f_{max}} \right)  \left( \frac{g}{\nu^2} \right),
 \label{eq:dx}
 \end{equation}
where $C$ is a dimensionless prefactor.  This equation includes the correct 
scaling of   ${\Delta x}_{max}$ with the frequency $\nu$ (Fig. \ref{fig17}(a)) 
and with the force $f_{max}$ (Fig. \ref{fig18}) . 
Interestingly, Eq. \ref{eq:dx} predicts that ${\Delta x}_{max}$ varies as $g^2$. 
This prediction is again in excellent agreement with our simulation data 
shown in Fig. \ref{fig19} for four different values of $g$.

\begin{figure}
\includegraphics[width=6cm]{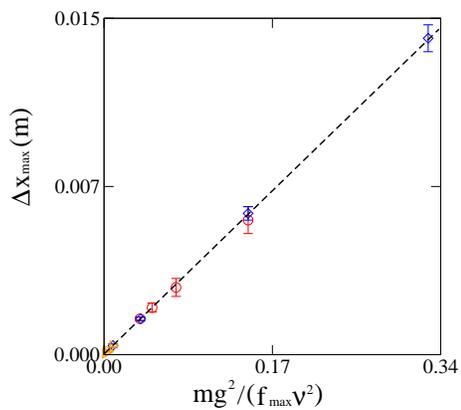}
\caption{Scaling of the maximum displacement ${\Delta x}_{max}$ with loading parameters 
from simulations with different values of the frequency $\nu$ (squares), the 
force amplitude $f_{max}$ (circles) and the gravity $g$ (diamonds).  \label{fig20}}
\end{figure}

Equation \ref{eq:dx} implies that the prefactor 
$C$ is a material constant that remains independent of 
all our loading parameters. Fig. \ref{fig20} shows  
${\Delta x}_{max}$ as a function of $mg^2/(f_{max} \nu^2)$ from 
 different simulations with different values of $\nu$, $f_{max}$ and $g$. 
 The data are in excellent agreement with the linear fit suggested by  Eq. \ref{eq:dx} with 
 $C \simeq 0.04$.

The above scaling can be incorporated in the fitting form \ref{eq:model} expressing 
$f_g$ as a function of $x-x_j$ and three forces $f_1$, $f_2$ and $f_3$; see Fig. \ref{fig9}.  
In this fitting form, the displacements should be divided by $\Delta x_{max} $. 
We will not study here in detail the dependence of $f_1$, $f_2$ and $f_3$ with respect to 
loading force parameters  $mg$, $f_{min}$ and $f_{max}$. Our simulations show 
that $f_3$ is independent of $f_{max}$, but it depends linearly on $mg$. Theoretically, 
this state corresponds to the limit active state where the ratio of principal stresses is 
a function of the internal angle of friction \cite{nedderman92}. On the other hand,
the force $f_2$ simply scales as $f_{max}$ and $f_1$ depends both on $f_{min}$ and $mg$. 
In our simulations, where $ f_{min} = 0$, the force $f_1$ is close to zero.  

%---Compaction rates

\section{Compaction rates}
\begin{figure}
\includegraphics[width=6cm]{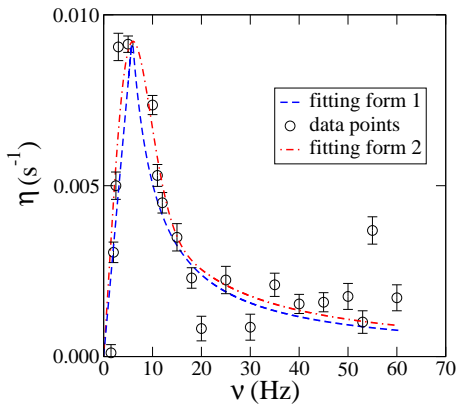}
\caption{The compaction rate $\eta$  as a function of  the frequency (circles) fitted by two 
different functions; see text. $\nu$\label{fig21}}
\end{figure}

Equation \ref{eq:eta1} suggests that the compaction rate $\eta$ should 
vary linearly with the frequency $\nu$ if the total compaction 
per period $\Delta \rho_1$ is independent of $\nu$. 
Fig. \ref{fig21} shows $\eta$  as a function of $\nu$. 
We see that only at low frequencies $\eta$ increases linearly with $\nu$. 
At larger frequencies, beyond a characteristic frequency $\nu_c$, 
$\eta$ declines with $\nu$. 
The largest compaction rate $\eta_{max}$ occurs for $\nu = \nu_c$. 
This implies that, according to Eq. \ref{eq:eta1},   
$\Delta \rho_1$ is indeed independent of $\nu$ for $\nu < \nu_c$. 
The characteristic time $\tau_c \equiv \nu_c^{-1}$ can be interpreted as the 
minimum time laps required for the relaxation of the packing. 
In fact, in the active state, the packing needs a finite rearrangement time $\tau_c$ 
to achieve a higher level of solid fraction. 
As long as the period $\tau = \nu^{-1}$ is longer than the relaxation time  
$\tau_c$, the packing has enough time  to relax fully to a more compact state. 
Then, the compaction $\Delta\rho_1$ has its maximum value $\Delta \rho_{max}$. 
But, if the  period $\tau$ is below  $\tau_c$, the relaxation will be incomplete so that 
 $\Delta \rho_1 < \Delta \rho_{max}$. 
 
Since  the volume change $\Delta V$ is proportional to $\Delta x$,  
$\Delta \rho_1$ follows the same scaling 
with the frequency as the displacement of 
the retaining wall, i.e.  $\Delta \rho_1 \propto \Delta \rho_{max} \ \nu^{-2}$. 
Hence, from Eq.  \ref{eq:eta1} and imposing the continuity at $\nu=\nu_c$, we get 
\begin{equation}
\eta=
\left\{
\begin{array}{lll}
\frac{\Delta\rho_{max}}{\rho_0} \ \nu & \nu < \nu_c, \\
 \frac{\Delta \rho_{max}}{\rho_0} \nu_c^2 \ \nu^{-1} & \nu > \nu_c. 
\end{array}
\right.
\label{eq:eta1-2}
\end{equation}
This form (labeled 1) is plotted in Fig. \ref{fig21} together with the data points. 
It is remarkable that, although $\nu_c$ is the only fitting parameter,  
the compaction rate $\eta$ is well 
adjusted by Eq. \ref{eq:eta1-2}. 
The prefactor $\Delta \rho_{max} / \rho_0$ is $ \simeq 1.5 \times 10^{-3}$,  
corresponding to   $\Delta \rho_{max} \simeq 1.1 \times 10^{-3}$. 

The arguments behind the proposed form \ref{eq:eta1-2} imply a 
sharp transition at $\nu = \nu_c$. 
This is rather plausible in view of the numerical data shown in Fig. \ref{fig21}. 
Nevertheless, it is convenient to construct a single expression 
containing the correct behavior both at low and high 
frequencies. The following fitting form provides a good approximation 
as shown also in  Fig. \ref{fig21} (fitting form 2):
\begin{equation}
\eta = \frac{\Delta\rho_{max}}{\rho_0} \frac{1 + e^{  - \left( \frac{\nu}{\nu_c} - 1  \right)^2 }}
{1 +  \left( \frac{\nu}{\nu_c}   \right)^2} \ \nu.
\label{eq:eta1-3}
\end{equation}     

We have $\nu_c \simeq 6$ Hz corresponding to a characteristic time $\tau_c = 0.17$ s. 
This time interval is long compared to single grain dynamics under gravity. 
For instance, the time required for a small-size grain in our samples 
to fall down a distance equal to its diameter is about $0.002$ s. 
Several observations show that collective rearrangements in granular 
media are often a slow process (\cite{Deboeuf2005}).    
Let us finally recall that our findings concern the short-time behavior 
($\Delta t < 1$ s). At longer times,  $\eta$ declines with time, 
but the scaling with frequency according to Eq. \ref{eq:eta1-2} is 
expected to hold at each instant of evolution of the packing.

% -----------------------------------------------------------------------------------
% Conclusion

\section{Conclusion}

In this paper, we analyzed the short-time behavior of a constrained granular system 
subjected to vibrational dynamics. 
The vibrations are induced by harmonic variation of the force exerted 
on a free retaining wall between zero and a maximum force. 
The system as a whole has a single degree of freedom 
represented by the horizontal position of the free wall. 
This system involves  a jammed state separating passive (loading) 
and active (unloading) states. 
The contact dynamics simulations were conducted with a rather small 
number of polygonal grains allowing for a systematic study of 
the dynamics and compaction of the material by varying the 
frequency and averaging over several configurations each time. 
By construction, our system is devoid of elastic elements and, hence, 
the behavior is fully governed by collective grain rearrangements. 

In the loading phase, the reaction force 
(exerted by  the grains on the free wall) rises almost linearly with the 
displacement of the  free wall, but it increases considerably at the end of this 
phase in transition to the jammed state. 
This force enhancement features the jamming transition 
compared to the rest of the passive state. 
The reaction force decreases then in the jammed state, balancing
 thus exactly the driving force, until the latter is low enough for the grains 
 to push the free wall away under the action of their own weights. 
 This unjamming process occurs smoothly and the reaction 
 force increases only slightly but exponentially during the unloading phase. 
We showed that a rough expression of the reaction force as a function 
of the displacement of the free wall with respect to the jamming 
position, provides a good prediction of the dynamics except at 
the jamming and unjamming transients. 

We used dimensional analysis to scale the dynamics with the frequency $\nu$ of oscillations. 
It was shown  that the data for frequencies ranging from 1 Hz to 60 Hz, collapse by scaling the displacements by the inverse square of frequency. On the other hand, we studies both 
numerically and dimensionally the scaling with loading parameters $mg$ and $f_{max}$.

We also investigated the oscillatory compaction of our numerical samples. 
A small compaction occurs during unloading, i.e. during the extension 
of the sample, followed by a smaller decompaction during loading. 
The compaction rate is nearly constant for short times. 
It was shown that the compaction rate increases linearly with frequency 
up to a characteristic frequency and then it declines nearly in inverse 
proportion to frequency. 
The characteristic frequency was interpreted in terms of the time 
required for the relaxation of a packing in each period to a more compact 
state by collective grain rearrangements under the action of gravity. 
The decreasing compaction rate as a function of frequency beyond 
the characteristic frequency was explained by arguing that 
only a partial relaxation, inversely proportional to frequency, could 
occur at such frequencies.  
  
A similar investigation is currently under way with polyhedral grains in three dimensions. 
Our preliminary results are consistent with those presented in this paper. 
In view of applications to a wider range of boundary conditions or diving modes, 
it is also important to consider in detail the characteristic time and the 
influence of various parameters pertaining to particle properties. 
Finally, long-time behavior and slow evolution of the compaction 
rate may be studied in this framework though more numerical effort is 
necessary to reach significant results in this case.    

%Acknowledgements

This work was funded by M. Valery from RFF (RŽ\'eseau Ferr\'eŽ Fran\c{c}ais) and 
the R\'egion Languedoc-Roussillon who are gratefully acknowledged.

% -----------------------------------------------------------------------------------
% Biblio


\begin{thebibliography}{27}
\expandafter\ifx\csname natexlab\endcsname\relax\def\natexlab#1{#1}\fi
\expandafter\ifx\csname bibnamefont\endcsname\relax
  \def\bibnamefont#1{#1}\fi
\expandafter\ifx\csname bibfnamefont\endcsname\relax
  \def\bibfnamefont#1{#1}\fi
\expandafter\ifx\csname citenamefont\endcsname\relax
  \def\citenamefont#1{#1}\fi
\expandafter\ifx\csname url\endcsname\relax
  \def\url#1{\texttt{#1}}\fi
\expandafter\ifx\csname urlprefix\endcsname\relax\def\urlprefix{URL }\fi
\providecommand{\bibinfo}[2]{#2}
\providecommand{\eprint}[2][]{\url{#2}}

\bibitem[{\citenamefont{Knight et~al.}(1995)\citenamefont{Knight, Fandrich,
  Lau, Jaeger, and Nagel}}]{Knight94}
\bibinfo{author}{\bibfnamefont{J.}~\bibnamefont{Knight}},
  \bibinfo{author}{\bibfnamefont{C.}~\bibnamefont{Fandrich}},
  \bibinfo{author}{\bibfnamefont{C.~N.} \bibnamefont{Lau}},
  \bibinfo{author}{\bibfnamefont{H.}~\bibnamefont{Jaeger}}, \bibnamefont{and}
  \bibinfo{author}{\bibfnamefont{S.}~\bibnamefont{Nagel}},
  \bibinfo{journal}{Phys. Rev. E.} \textbf{\bibinfo{volume}{51}},
  \bibinfo{pages}{3957} (\bibinfo{year}{1995}).

\bibitem[{\citenamefont{Sano}(2005)}]{Sano2005}
\bibinfo{author}{\bibfnamefont{O.}~\bibnamefont{Sano}}, \bibinfo{journal}{Phys.
  Rev. E.} \textbf{\bibinfo{volume}{72}}, \bibinfo{pages}{3957}
  (\bibinfo{year}{2005}).

\bibitem[{\citenamefont{Aoki et~al.}(1996)\citenamefont{Aoki, Akiyama, Maki,
  and Watanabe}}]{aoki96}
\bibinfo{author}{\bibfnamefont{K.~M.} \bibnamefont{Aoki}},
  \bibinfo{author}{\bibfnamefont{T.}~\bibnamefont{Akiyama}},
  \bibinfo{author}{\bibfnamefont{Y.}~\bibnamefont{Maki}}, \bibnamefont{and}
  \bibinfo{author}{\bibfnamefont{T.}~\bibnamefont{Watanabe}},
  \bibinfo{journal}{Phys. Rev. E} \textbf{\bibinfo{volume}{54}},
  \bibinfo{pages}{874} (\bibinfo{year}{1996}).

\bibitem[{\citenamefont{Liffman et~al.}(1997)\citenamefont{Liffman, Metcalfe,
  and Cleary}}]{liffman97}
\bibinfo{author}{\bibfnamefont{K.}~\bibnamefont{Liffman}},
  \bibinfo{author}{\bibfnamefont{G.}~\bibnamefont{Metcalfe}}, \bibnamefont{and}
  \bibinfo{author}{\bibfnamefont{P.}~\bibnamefont{Cleary}},
  \bibinfo{journal}{Phys. Rev. Lett.} \textbf{\bibinfo{volume}{79}},
  \bibinfo{pages}{4574} (\bibinfo{year}{1997}).

\bibitem[{\citenamefont{Knight et~al.}(1993)\citenamefont{Knight, Jaeger, and
  Nagel}}]{Knight93}
\bibinfo{author}{\bibfnamefont{J.~B.} \bibnamefont{Knight}},
  \bibinfo{author}{\bibfnamefont{H.~M.} \bibnamefont{Jaeger}},
  \bibnamefont{and} \bibinfo{author}{\bibfnamefont{S.~R.} \bibnamefont{Nagel}},
  \bibinfo{journal}{Phys. Rev. E} \textbf{\bibinfo{volume}{74}},
  \bibinfo{pages}{3728} (\bibinfo{year}{1993}).

\bibitem[{\citenamefont{Aoki and Akiyama}(1996)}]{aoki96b}
\bibinfo{author}{\bibfnamefont{K.~M.} \bibnamefont{Aoki}} \bibnamefont{and}
  \bibinfo{author}{\bibfnamefont{T.}~\bibnamefont{Akiyama}},
  \bibinfo{journal}{Phys. Rev. Lett.} \textbf{\bibinfo{volume}{77}},
  \bibinfo{pages}{4166} (\bibinfo{year}{1996}).

\bibitem[{\citenamefont{Clement et~al.}(1996)\citenamefont{Clement, Vanel,
  Rajchenbach, and J.Duran}}]{Clement96}
\bibinfo{author}{\bibfnamefont{E.}~\bibnamefont{Clement}},
  \bibinfo{author}{\bibfnamefont{L.}~\bibnamefont{Vanel}},
  \bibinfo{author}{\bibfnamefont{J.}~\bibnamefont{Rajchenbach}},
  \bibnamefont{and} \bibinfo{author}{\bibnamefont{J.Duran}},
  \bibinfo{journal}{Phys. Rev. E} \textbf{\bibinfo{volume}{53}},
  \bibinfo{pages}{2972} (\bibinfo{year}{1996}).

\bibitem[{\citenamefont{Jaeger et~al.}(1996)\citenamefont{Jaeger, Nagel, and
  Behringer}}]{jaeger96b}
\bibinfo{author}{\bibfnamefont{H.~M.} \bibnamefont{Jaeger}},
  \bibinfo{author}{\bibfnamefont{S.~R.} \bibnamefont{Nagel}}, \bibnamefont{and}
  \bibinfo{author}{\bibfnamefont{R.~P.} \bibnamefont{Behringer}},
  \bibinfo{journal}{Reviews of Modern Physics} \textbf{\bibinfo{volume}{68}},
  \bibinfo{pages}{1259} (\bibinfo{year}{1996}).

\bibitem[{\citenamefont{Brennen et~al.}(1993)\citenamefont{Brennen, Ghosh, and
  Wassgren}}]{Brennen93}
\bibinfo{author}{\bibfnamefont{C.}~\bibnamefont{Brennen}},
  \bibinfo{author}{\bibfnamefont{S.}~\bibnamefont{Ghosh}}, \bibnamefont{and}
  \bibinfo{author}{\bibfnamefont{C.}~\bibnamefont{Wassgren}}, in
  \emph{\bibinfo{booktitle}{Powders and Grains 93}} (\bibinfo{publisher}{A. A.
  Balkema}, \bibinfo{address}{Amsterdam}, \bibinfo{year}{1993}), pp.
  \bibinfo{pages}{247--252}.

\bibitem[{\citenamefont{Weathers et~al.}(1997)\citenamefont{Weathers, Hunt,
  Brennen, Lee, and Wassgren}}]{weathers97}
\bibinfo{author}{\bibfnamefont{R.~C.} \bibnamefont{Weathers}},
  \bibinfo{author}{\bibfnamefont{M.~L.} \bibnamefont{Hunt}},
  \bibinfo{author}{\bibfnamefont{C.~E.} \bibnamefont{Brennen}},
  \bibinfo{author}{\bibfnamefont{A.~T.} \bibnamefont{Lee}}, \bibnamefont{and}
  \bibinfo{author}{\bibfnamefont{C.~R.} \bibnamefont{Wassgren}},
  \emph{\bibinfo{title}{Effects of horizontal vibration on hopper flows of
  granular material}} (\bibinfo{year}{1997}), pp. \bibinfo{pages}{349--360}.

\bibitem[{\citenamefont{Wassgren et~al.}(1997)\citenamefont{Wassgren, Hunt, and
  Brennen}}]{wassgren97c}
\bibinfo{author}{\bibfnamefont{C.~R.} \bibnamefont{Wassgren}},
  \bibinfo{author}{\bibfnamefont{M.~L.} \bibnamefont{Hunt}}, \bibnamefont{and}
  \bibinfo{author}{\bibfnamefont{C.~E.} \bibnamefont{Brennen}},
  \emph{\bibinfo{title}{Effects of vertical vibration on hopper flows of
  granular material}} (\bibinfo{year}{1997}), pp. \bibinfo{pages}{335--348}.

\bibitem[{\citenamefont{Saussine}(october 2004)}]{SAUSSINE2004}
\bibinfo{author}{\bibfnamefont{G.}~\bibnamefont{Saussine}}, Ph.D. thesis,
  \bibinfo{school}{UniversitÈ Montpellier II} (\bibinfo{year}{october 2004}).

\bibitem[{\citenamefont{Oviedo}(May 2001)}]{XIMENA2001}
\bibinfo{author}{\bibfnamefont{X.}~\bibnamefont{Oviedo}}, Ph.D. thesis,
  \bibinfo{school}{LCPC} (\bibinfo{year}{May 2001}).

\bibitem[{\citenamefont{Markland}(1981)}]{MORGAN81}
\bibinfo{author}{\bibfnamefont{J.~M.~E.} \bibnamefont{Markland}},
  \bibinfo{journal}{Geotechnique} \textbf{\bibinfo{volume}{31}},
  \bibinfo{pages}{3,367} (\bibinfo{year}{1981}).

\bibitem[{\citenamefont{Luding}(95)}]{Luding95}
\bibinfo{author}{\bibfnamefont{S.}~\bibnamefont{Luding}},
  \bibinfo{journal}{Phys. Rev. E} \textbf{\bibinfo{volume}{52}},
  \bibinfo{pages}{52 } (\bibinfo{year}{95}).

\bibitem[{\citenamefont{Ben-Naim et~al.}(1996)\citenamefont{Ben-Naim, Knight,
  and Nowak}}]{ben-naim96}
\bibinfo{author}{\bibfnamefont{E.}~\bibnamefont{Ben-Naim}},
  \bibinfo{author}{\bibfnamefont{J.~B.} \bibnamefont{Knight}},
  \bibnamefont{and} \bibinfo{author}{\bibfnamefont{E.~R.} \bibnamefont{Nowak}},
  \bibinfo{journal}{J. Chem. Phys.} \textbf{\bibinfo{volume}{100}},
  \bibinfo{pages}{6778} (\bibinfo{year}{1996}).

\bibitem[{\citenamefont{Ben-Naim et~al.}(1997)\citenamefont{Ben-Naim, Knight,
  Nowak, Jaeger, and Nagel}}]{ben-naim97}
\bibinfo{author}{\bibfnamefont{E.}~\bibnamefont{Ben-Naim}},
  \bibinfo{author}{\bibfnamefont{J.~B.} \bibnamefont{Knight}},
  \bibinfo{author}{\bibfnamefont{E.~R.} \bibnamefont{Nowak}},
  \bibinfo{author}{\bibfnamefont{H.~M.} \bibnamefont{Jaeger}},
  \bibnamefont{and} \bibinfo{author}{\bibfnamefont{S.~R.} \bibnamefont{Nagel}}
  (\bibinfo{year}{1997}), \bibinfo{note}{submitted to the proceedings of the
  17th annual CNLS conference ``nonlinear waves in physical phenomena''}.

\bibitem[{\citenamefont{Hunt et~al.}(1999)\citenamefont{Hunt, Weathers, Lee,
  and Brennen}}]{Hunt99}
\bibinfo{author}{\bibfnamefont{M.~L.} \bibnamefont{Hunt}},
  \bibinfo{author}{\bibfnamefont{R.~C.} \bibnamefont{Weathers}},
  \bibinfo{author}{\bibfnamefont{A.~T.} \bibnamefont{Lee}}, \bibnamefont{and}
  \bibinfo{author}{\bibfnamefont{C.~E.} \bibnamefont{Brennen}},
  \bibinfo{journal}{Phys. Rev. E} \textbf{\bibinfo{volume}{11}},
  \bibinfo{pages}{68 } (\bibinfo{year}{1999}).

\bibitem[{\citenamefont{Kudrolli}(2004)}]{Kudrolli2004}
\bibinfo{author}{\bibfnamefont{A.}~\bibnamefont{Kudrolli}},
  \bibinfo{journal}{Rep. Prog. Phys} \textbf{\bibinfo{volume}{67}},
  \bibinfo{pages}{209} (\bibinfo{year}{2004}).

\bibitem[{\citenamefont{Josserand et~al.}(2000)\citenamefont{Josserand,
  Tkachenko, Mueth, and Jaeger}}]{Josserand2000}
\bibinfo{author}{\bibfnamefont{C.}~\bibnamefont{Josserand}},
  \bibinfo{author}{\bibfnamefont{A.~V.} \bibnamefont{Tkachenko}},
  \bibinfo{author}{\bibfnamefont{D.~M.} \bibnamefont{Mueth}}, \bibnamefont{and}
  \bibinfo{author}{\bibfnamefont{H.~M.} \bibnamefont{Jaeger}},
  \bibinfo{journal}{Phys. Rev. E} \textbf{\bibinfo{volume}{85}},
  \bibinfo{pages}{3632 } (\bibinfo{year}{2000}).

\bibitem[{\citenamefont{Jean and Moreau}(1992)}]{jean92}
\bibinfo{author}{\bibfnamefont{M.}~\bibnamefont{Jean}} \bibnamefont{and}
  \bibinfo{author}{\bibfnamefont{J.~J.} \bibnamefont{Moreau}}, in
  \emph{\bibinfo{booktitle}{Proceedings of Contact Mechanics International
  Symposium}} (\bibinfo{publisher}{Presses Polytechniques et Universitaires
  Romandes}, \bibinfo{address}{Lausanne, Switzerland}, \bibinfo{year}{1992}),
  pp. \bibinfo{pages}{31--48}.

\bibitem[{\citenamefont{Moreau}(2004)}]{mo04a}
\bibinfo{author}{\bibfnamefont{J.}~\bibnamefont{Moreau}}, in
  \emph{\bibinfo{booktitle}{Novel approaches in civil engineering}}, edited by
  \bibinfo{editor}{\bibfnamefont{M.}~\bibnamefont{Fr{\'e}mond}}
  \bibnamefont{and} \bibinfo{editor}{\bibfnamefont{F.}~\bibnamefont{Maceri}}
  (\bibinfo{publisher}{Springer-Verlag}, \bibinfo{year}{2004}),
  no.~\bibinfo{number}{14} in \bibinfo{series}{Lecture Notes in Applied and
  Computational Mechanics}, pp. \bibinfo{pages}{1--46}.

\bibitem[{\citenamefont{Dubois and Jean}(volume1, CSMA-AFM-LMS,
  2003)}]{DUBOIS2003}
\bibinfo{author}{\bibfnamefont{F.}~\bibnamefont{Dubois}} \bibnamefont{and}
  \bibinfo{author}{\bibfnamefont{M.}~\bibnamefont{Jean}},
  \bibinfo{journal}{Actes du sixi\`eme colloque national en calcul des
  structures}  (\bibinfo{year}{volume1, CSMA-AFM-LMS, 2003}).

\bibitem[{\citenamefont{Nedderman}(1992)}]{nedderman92}
\bibinfo{author}{\bibfnamefont{R.~M.} \bibnamefont{Nedderman}},
  \emph{\bibinfo{title}{Statics and kinematics of granular materials}}
  (\bibinfo{publisher}{Cambr. Univ. Press}, \bibinfo{address}{Cambridge},
  \bibinfo{year}{1992}).

\bibitem[{\citenamefont{GDRMiDi}(2004)}]{GDRMIDI2004}
\bibinfo{author}{\bibnamefont{GDRMiDi}}, \bibinfo{journal}{Eur. Phys. Rev. E}
  \textbf{\bibinfo{volume}{14}}, \bibinfo{pages}{341} (\bibinfo{year}{2004}).

\bibitem[{\citenamefont{Mitchell and Soga}(2005)}]{mitchell05}
\bibinfo{author}{\bibfnamefont{J.}~\bibnamefont{Mitchell}} \bibnamefont{and}
  \bibinfo{author}{\bibfnamefont{K.}~\bibnamefont{Soga}},
  \emph{\bibinfo{title}{Fundamentals of Soil Behavior}}
  (\bibinfo{publisher}{Wiley}, \bibinfo{address}{New York,{USA}},
  \bibinfo{year}{2005}).

\bibitem[{\citenamefont{Deboeuf et~al.}(2005)\citenamefont{Deboeuf, O.Dauchot,
  Staron, Mangeney, and Vilotte}}]{Deboeuf2005}
\bibinfo{author}{\bibfnamefont{S.}~\bibnamefont{Deboeuf}},
  \bibinfo{author}{\bibnamefont{O.Dauchot}},
  \bibinfo{author}{\bibfnamefont{L.}~\bibnamefont{Staron}},
  \bibinfo{author}{\bibfnamefont{A.}~\bibnamefont{Mangeney}}, \bibnamefont{and}
  \bibinfo{author}{\bibfnamefont{J.-P.} \bibnamefont{Vilotte}},
  \bibinfo{journal}{Phys. Rev. E} \textbf{\bibinfo{volume}{72}},
  \bibinfo{pages}{1} (\bibinfo{year}{2005}).

\end{thebibliography}
\end{document}